\documentclass[runningheads]{llncs}
\usepackage{tabularx}
\usepackage{graphicx}
\usepackage{xcolor}
\usepackage{tabu}
\usepackage{amssymb}
\usepackage{amsmath}
\newcommand{\probP}{\text{I\kern-0.15em P}}
\usepackage{url}

\begin{document}
\title{A causal model of safety assurance for machine learning}
%
%

\author{Simon Burton}

\authorrunning{S. Burton}
%
\institute{Fraunhofer Institute for Cognitive Systems, Munich, Germany
\email{simon.burton@iks.fraunhofer.de}}
\maketitle              
\begin{abstract}
 This paper proposes a framework based on a causal model of safety upon which effective safety assurance cases for ML-based applications can be developed. In doing so, we build upon established principles of safety engineering as well as previous work on structuring assurance arguments for ML. The paper defines four categories of safety case evidence and a structured analysis approach within which these evidences can be effectively combined. Where appropriate, abstract formalisations of these contributions are used to illustrate the causalities they evaluate, their contributions to the safety argument and desirable properties of the evidences. Based on the proposed framework, progress in this area is re-evaluated and a set of future research directions proposed in order for tangible progress in this field to be made.

\keywords{Machine learning  \and Safety of the intended functionality \and Safety assurance \and Cyber-physical systems}
\end{abstract}
\section{Introduction}
Innovations in machine learning (ML) have demonstrated their potential for performing tasks that were otherwise infeasible using software programming approaches. These advancements have unlocked the promise of greater levels of autonomy in various cyber-physical systems including automated driving systems, industrial robotics and driverless trains. Despite the potential for increasing safety by reducing the impact of human errors or by performing tasks otherwise deemed too hazardous for humans to perform, these systems also introduce new classes of risk. These are inherent in the transfer of complex, critical decisions from human operators to the system as well as the unpredictability of the operating domain \cite{burton2020mind}. The use of ML for perception and decision tasks within this context introduces new concerns and specific challenges for safety assurance. These include the use of training data in place of a detailed specification of the required (safe) behaviour, uncertainty within the results of the trained models and the lack of explainability in the calculations.

Standards related to the safety of electrical/electronic systems define functional safety as the absence of unreasonable risk due to hazards caused by malfunctioning behaviour of the systems typically interpreted as either random hardware faults or systematic errors in the system, hardware, or software. 
 
In addition to demonstrating adherence to systematically derived functional requirements, software safety assurance methods also target non-functional properties such as errors caused by pointer mismatches, division-by-zero, etc. Whilst the set of detailed methods for achieving safe software may be applied to the training and execution platforms of the ML models, they are not well suited to preventing or detecting errors caused by performance limitations (insufficiencies) within the trained model itself. For example, caused by a lack of suitable training data or unforeseen changes in the operating environment. 

The standard ISO 21448 ``Road vehicles - Safety of the intended functionality (SOTIF)'' addresses safety in terms of the absence of unreasonable risk due to functional insufficiencies of the system or by reasonably foreseeable misuse. The SOTIF model therefore seems well suited for arguing the performance of the trained ML model over all possible scenarios within a given operating domain, where \textit{insufficiencies}, are uncovered by certain \textit{triggering conditions} in the inputs that could lead to hazards. The standard requires the definition of quantitative \textit{acceptance criteria} and related \textit{validation targets} for each safety goal, which in turn can be refined and allocated to subsystems such as perception or decision functions. 

This paper proposes a set of safety argumentation strategies that could form the basis of an assurance case for the safety-related performance of ML-based applications. Wherever possible, established principles of safety engineering will be applied, but adapted to the specific challenges of machine learning safety. These include the consideration of causal relationships between failures and their causes, the application of structured safety analyses to identify and evaluate appropriate safety measures and the collection of a diverse set of evidences for use within a safety assurance case. The concepts described within this paper are developed with supervised machine learning paradigms in mind. However, it is expected that the principles could be adapted to the safety-related application of other forms to data-intensive functions including reinforcement learning and statistical modelling. 

The paper is structured as follows. Section \ref{Sec_Related_Work} summarises related work in this area upon which the framework described here builds. In Section \ref{Sec_Causal_Model}, a causal model of safety-related failures of ML-based applications is introduced. This is developed in Section \ref{Sec_Categories} to illustrate the argumentation strategies that form part of an iterative cycle of safety assurance. Where appropriate, the concepts will be supported by an abstract formalisation to illustrate the causalities under investigation and that can be used to reason about the effectiveness of concrete methods and metrics specific to particular applications and ML-approaches. The paper concludes with a reflection on current progress in this field of research and an agenda for future work.

\section{Related work and preliminaries}\label{Sec_Related_Work}
Previous work on the safety assurance of machine learning has focused on the structure of the assurance case and associated processes with respect to existing safety standards \cite{Burton2017making,salay2017analysis,gauerhof2020assuring,ashmore2021assuring,burton2021safety}. Other work has focused on the effectiveness of specific metrics and measures on providing meaningful statements related to safety properties of the ML function \cite{cheng2018towards,henne2020benchmarking,Schwaiger2021,cheng2021safety}. This paper therefore complements these two strands of work by examining in more detail the role of combining various types of evidences to form a convincing argument that quantitative acceptance criteria as required by standards such as ISO 21448 are met. In doing so, we extend the causal model which was first informally introduced in \cite{BookBurton2022}.

We use the term \textit{ML System} ($M_S$) to denote the scope of the safety assurance activities which includes the training data, architecture, hyper-parameters, the trained model itself ($M_M$) and post-processing functions (see Figure \ref{Fig_Causal_Model}). The starting point of the safety assurance activities must be some definition of acceptably safe behaviour of the ML system in terms of a set of technical safety requirements. A complete specification of the correct outputs for each possible input is not feasible for all but the most trivial applications of machine learning. Therefore, the safety properties of the ML system must be expressed as a set of constraints on the behaviour of the model, that may allow for some bounded deviations from the ground truth for individual inputs. 
In \cite{Burton2017making}, the authors recommend contract-based design to define such safety requirements. These requirements are expressed in the form of a set of guarantees that the function must fulfil whenever the inputs satisfy a set of assumptions on the system's environmental and technical context in order for system safety goals to be met. This requires a suitable definition of safety at the level of abstraction of the ML system and the model itself. For example, a superficially defined requirement, such as ``detect a pedestrian in the road", would need to be refined to define which characteristics constitute a pedestrian, from which distance and level of occlusion pedestrians should be detected, and the number of consecutive mis-classifications that could be tolerated before the system performs an unsafe control action. Additional safety arguments are required that the set of assumptions is valid and maintained during operation, however we do not address this issue in more detail here. The specification of such properties (safety guarantees) has been explored by a number of authors \cite{bergenhem2015reach,gauerhof2018structuring,hu2020towards,ashmore2021assuring} and can include the following categories:

\begin{itemize}
    \item \textit{Functional properties} of the task, e.g. definition of the objects to be identified and range of sizes and distances of relevant objects.
    \item \textit{Accuracy and precision} with which the task should be performed. This can include bounds on the rate of incorrect outputs of the model, e.g. number of false positive or false negative results and the amount of deviation from ground truth. This may include target average failure rates but also constraints on the distribution of failures such as a maximum number of successive errors or equal distribution of errors across input classes. 
    \item \textit{Robustness} against small changes in the input space that occur either naturally within the operating domain (e.g. lighting conditions) or technical system (e.g. sensor noise). This can also include sensitivity to intentionally adversarial examples \cite{goodfellow2014explaining}.
    \item \textit{Prediction certainty} provided by the model. For example in the soft-max values provided by a neural network,  the confidence value provided by the model should correspond to the actual probability of error. For example, for all objects detected by the ML model with a confidence score of 0.8, the probability of a correct detection should be approximately 80\%.
\end{itemize}

For each safety goal at the system level, a different set of guarantees, or parameters thereof may be required. An argument that the safety requirements on the ML system are free from significant specification insufficiencies is an essential part of safety assurance activities, but will not be elaborated in detail in this paper. In \cite{burton2019confidence}, the concept of the safety contract for an ML system was expressed as the following condition that must be fulfilled for the ML system to be considered safe.

\begin{equation}\label{Eq_Safety_Contract}
	\forall i \in I.A(i) \Rightarrow G(i, M_S(i))
\end{equation}

Informally, for all inputs $i$ that fulfil the set of assumptions $A$ on the operating domain and system context, the output of the model must fulfill a set of conditions defined by the guarantees $G$. Practically, it will not be possible to ``prove'' equation \ref{Eq_Safety_Contract} for all possible inputs. Therefore the condition will need to be \textit{inferred} in an inductive manner based on evidence that is collected about the design and performance of the ML system, which is the inherent nature of most forms of safety assurance. This leads to the the concept of a quantitative acceptance criteria as proposed by ISO 21448, where the ML system can be considered ``acceptably safe" under the following conditions.



Precisely, let $\probP_{ODD}: I \rightarrow [0,1]$ be the \emph{input probability distribution function} of the ODD that assigns every input $i \in I$ with a probability value, with the condition that $\sum_{i \in I} \probP_{ODD}(i) = 1$. For an ML system to be acceptably safe, the following condition should hold:

\begin{equation}\label{Eq_Theory}
 \frac{ \sum_{i \in I, A(i) \wedge G(i, M_S(i))} \probP_{ODD}(i)}{ \sum_{i \in I, A(i)}  \probP_{ODD}(i)}  	 \geq AC
\end{equation}

In Equation~\ref{Eq_Theory}, the left-hand side characterizes the \emph{conditional probability} of an input satisfying the guarantee $G$, conditional to the constraint that assumption $A$ holds. Equation~\ref{Eq_Theory} essentially states that so long as the failure rate (where the probability of ($G(i, M_S(i)) = false$) is small, the system is considered to be acceptably safe. The value $AC$ (acceptance criteria) on the right-hand side is the minimum tolerable level of probability of the model fulfilling its guarantees. 

Unfortunately,  $\probP_{ODD}$ acting as the input distribution function can never be perfectly characterized for complex systems such as autonomous driving. This highlights one of the challenges in calculating realistic failures rates for such systems, as any measurements will ultimately be sensitive to the unknown distribution of critical and non-critical events (triggering conditions) in the input domain. Any definition or measurement of failure rates for such systems will therefore only ever be an approximation of the actual failure rates experienced during operation.

Given that the conditions given in Equation~\ref{Eq_Safety_Contract} cannot be proven with absolute certainty for realistic systems, the challenge therefore is to find a set of conditions that can be demonstrated with sufficient confidence from which we can infer that the condition set out in Equation~\ref{Eq_Contract_Performance} is met. This leads to the concept of \emph{measured failure rate} that serves as a proxy for Equation~\ref{Eq_Theory} when constructing assurance arguments about the system.

\begin{equation}\label{Eq_PFOD}
 	 \lambda_{MS} = \frac{\#\{i \in I : A(i) \land \neg G(i, M_{S}(i))\}}
 	{\#\{i \in I : A(i)\}}
\end{equation}

\begin{equation}\label{Eq_Contract_Performance}
 	 1 - \lambda_{MS}\geq AC
\end{equation}

Where $I$ is now used to represent the \emph{measured} input space of unique samples over time, rather than the entire input space that could be theoretically experienced during operation. Here the $\lambda_{MS}$ serves as a proxy for the \emph{measured probability of failure on demand} under the assumption that all inputs in the domain may occur with equal probability. We build upon these definitions later in the paper whilst describing the contribution of different categories of evidence to arguing that these conditions are met. This simplification has consequences in the choice of inputs used to create safety evidence. Inputs should be chosen to reflect the expected distribution in the target ODD. 

\section{A causal model of machine learning safety}\label{Sec_Causal_Model}

\begin{figure*}[ht]
    \centering
    \includegraphics[width=\textwidth]{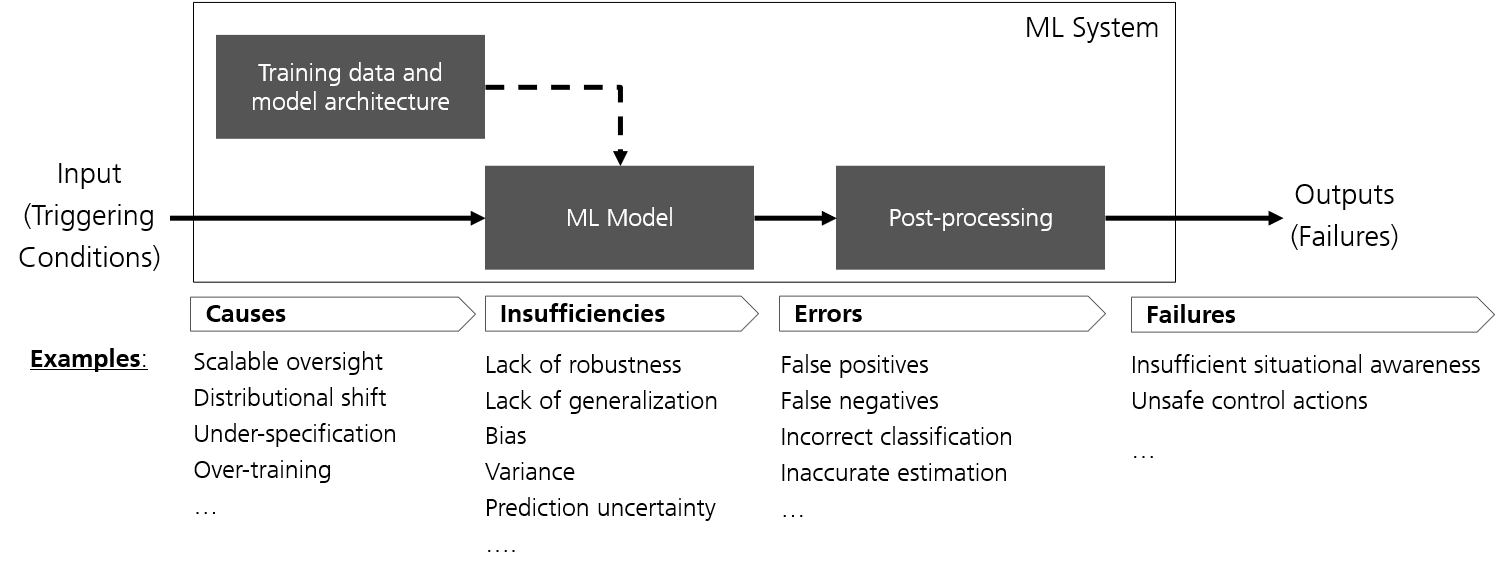}
    \caption{Causal model of safety-relevant ML failures}
    \label{Fig_Causal_Model}
\end{figure*}

The dependability model of Avizienis et al.~\cite{avizienis2004basic} describes the foundation of many safety analysis techniques. In this model, the risk of a system violating its objectives is determined by analysing the propagation path from causes of individual \textit{faults} that could lead to an \textit{erroneous} state which in turn leads to a \textit{failure} to maintain the system's objectives. Risk can thus be controlled by either eliminating the causes of faults or by preventing them from leading to potentially hazardous erroneous states of the system. Figure \ref{Fig_Causal_Model} summarises how this causal model can be applied to the problem of assessing and ensuring the safety-related performance associated with insufficiencies in the trained ML model, here using examples related to the use of Deep Neural Networks (DNN) for safety-relevant functions. A similar set of causal relations can be considered where errors are triggered due to faults in the execution of the model itself, caused by software or hardware faults in the execution platform. These classes of safety-related failures would be handled by functional safety approaches as defined by existing standards such as IEC 61508 or ISO 26262 and are not considered in detail here.  

The causal model illustrated in this paper is based on an architecture associated with supervised machine learning, where a pre-trained model $M_M$ is presented with novel inputs during operation and where a post-processing step is used to prevent, detect or compensate for errors and inaccuracies in the model. Examples of commonly used post-processing approaches are the suppression of model outputs for out-of-distribution inputs, redundant calculations or plausibility checks. Within this context, a safety-related failure can be defined as a condition where an erroneous output of $M_M$ is insufficiently compensated for by the post-processing step such that it can lead to an unsafe action of the ML system $M_S$. Erroneous model outputs are caused by insufficiencies in the ML model, uncovered by certain inputs (triggering conditions). These insufficiencies, in turn, could be caused by limitations in the training data or the model architecture itself. Examples of such causes include lack of scalable oversight \cite{amodei2016concrete}, ontological uncertainty \cite{gansch2020system} (insufficient knowledge about the input domain) and distributional shift \cite{amodei2016concrete}.

\begin{figure*}[ht]
    \centering
    \includegraphics[width=0.7\textwidth]{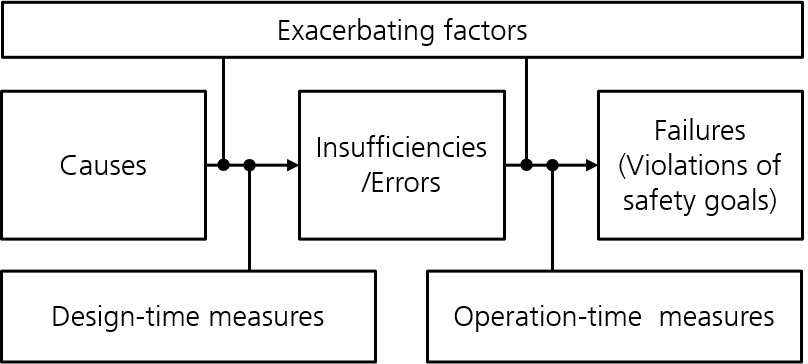}
    \caption{Summary of safety analysis approach}
    \label{Fig_Analysis_Model}
\end{figure*}

This causal model can be used as the basis for analysis techniques for identifying and evaluating the effectiveness of ML safety measures. Measures to improve the safety of the ML system can be categorised into those applied during design time or during operation.  The underlying structure to these safety analyses is summarised in Figure \ref{Fig_Analysis_Model} which is loosely based on the bow-tie diagram method of safety analysis \cite{de2016bowtie}. Inductive analyses similar to Failure Modes and Effects Analysis (FMEA) \cite{mcdermott1996basics} would first consider the various types of insufficiencies (failure modes) that could lead to the violation of a safety goal and their underlying causes. Design-time measures would be evaluated regarding their effectiveness at suppressing the causes that lead to the insufficiencies and measuring resulting properties of the model. Operation-time measures would be evaluated at their effectiveness to prevent residual errors from leading to a violation of the safety goals. During such analysis it is also important to consider exacerbating factors that could minimise the effectiveness of the safety measures. These could include for example the lack of availability of certain classes of training data to counteract bias in the model and the lack of explainability of the model results which limit the potential of post-processing techniques or the causal analysis itself. Deductive forms of safety analysis could include evaluating specific safety-related failures discovered during validation to determine the underlying causes and improvements to the ML system required to prevent their occurrence in future.

\begin{figure*}[ht]
   \centering
    \includegraphics[width=0.95\textwidth]{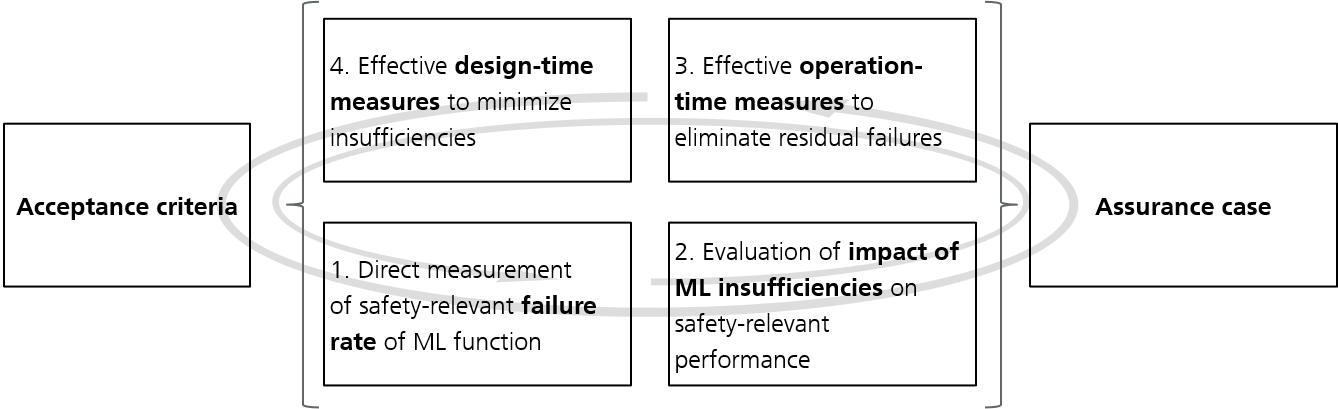}
    \caption{Safety assurance framework for machine learning systems }
    \label{Fig_Assurance_Framework}
\end{figure*}

As each ML model may need to fulfill multiple safety goals leading to potentially conflicting acceptance criteria, safety assurance of ML systems can be considered as an optimisation problem. As illustrated in Figure \ref{Fig_Assurance_Framework}, an iterative approach is proposed to converge to this optimum where each cycle begins with a set of acceptance criteria defined in the form of a safety contract (see Section \ref{Sec_Related_Work}) that includes defined bounds on errors and uncertainties within the model. As will become clear in the following sections, the categories of evidence interact and a suitable combination of methods will support a more convincing safety assurance argument. The safety assurance cycle is repeated until a sufficiently convincing argument can be constructed based on the fulfilment of all relevant quantitative and qualitative acceptance criteria. It is also to be expected that for many applications, such an argument may not be possible without further restricting the expectations on the ML systems including, for example, the scope of relevant operational scenarios and requirements on the availability of the function.

In the following section, each category of evidence will be described in more detail, supported where appropriate by abstract formalisms to demonstrate desirable properties that would lead to a convincing assurance argument. The order in which these categories of evidence are presented here assumes that the safety activities begin with a trained model that has reached a certain level of baseline performance. However, due to the iterative nature of the approach and the need to combine various types of complementary evidence, the order in which the evidences are collected in practice is also somewhat arbitrary.

\section{Categories of safety evidence for machine learning}\label{Sec_Categories}

\subsection{Category 1: Direct measurement of the failure rate of the ML system} \label{Sub_Failure_rates}
Equation \ref{Eq_Contract_Performance} was used to define the condition that the ML system meets the acceptance criteria for a given safety goal by comparing the probability that the safety contract has been fulfilled with a quantitative definition of the acceptance criteria. ISO 21448 introduces the concept of \textit{validation targets} which are values that can be used to argue that the acceptance criteria are met. For any given safety goal and acceptance criteria, a combination of validation targets may be required in order to provide a convincing argument. Validation targets can be assessed by evaluating a sample set $S$ of the entire input space $I$, limited to those inputs that fulfill the assumptions $A$. This leads to the following generic definition of the condition for meeting a validation target $VT$, where $E$ is an \textit{evidence function} that returns some value based on the sample set $S$ and the model $M_S$ (also derived from \cite{burton2019confidence}).

\begin{equation}\label{Eq_Validation}
 \begin{array}{c}
 	S \subseteq I \land \forall j \in S.A(j) \Rightarrow E(S, M_S) \geq VT 
 \end{array}
\end{equation}

Another way of understanding such evidence functions and associated validation targets are to treat them as the ``known facts'' upon which we will infer whether the acceptance criteria for the system have been fulfilled, as expressed in equation \ref{Eq_Contract_Performance}. In its most simple form, $E$ could be based on black-box testing techniques and provide an estimation of whether the quantitative acceptance criteria are met based on the pass rate of a set of representative tests $S$. It will not be possible to sample the entire input space $I$ or even $\{i \in I : A(i)\}$ due to the complexity and number of scenarios for most safety-relevant applications. If failures were equiprobable for each $i \in I$, and a sufficiently large portion of $I$ was sampled these tests would provide a suitable estimation of the actual probability of failure $\lambda_{MS}$ of the ML system. Unfortunately, this is not the case given what we know about the causes and manifestations of insufficiencies in machine learning models. Therefore, criteria for selecting the test inputs $S$ must ensure that sufficient test samples are chosen that are both representative of the input space and are capable of detecting failures in the function. For example, test selection criteria such as equivalence class coverage and testing edge cases \cite{hutchison2018robustness} improve the suitability of $S$. 

Statistically, the estimation of $\lambda_{MS}$ must have a variance commensurate to the target value of $\lambda_{MS}$. In other words, the lower the target value of $\lambda_{MS}$, the more exact the method of estimating $\lambda_{MS}$ must be. In order to increase confidence in the estimation, a better understanding of the causes, types and distribution of errors that lead to hazardous failures is required (see category 2, Section \ref{Sub_Insufficiencies}). 
The ability to extrapolate estimations of the failure rates of the ML \textit{system} based on observations of the errors at the outputs of the ML \textit{model} can be further enhanced based on an understanding of the effectiveness of the operation-time measures (see category 3, Section \ref{Sub_Operation_time}) at preventing errors of the ML model to propagate to safety-relevant failures. 

\subsection{Category 2: Evaluating the impact of insufficiencies in the ML model} \label{Sub_Insufficiencies}
Black-box testing approaches at the system level can be used to uncover previously \textit{unknown, unsafe triggering conditions}, whereby the strength of the argument regarding the residual risk associated with such conditions is limited by both the complexity of the operating domain and inherent properties of the ML model. Evidence of category 1 will typically therefore not be sufficient to infer that the acceptance criteria for the ML system have been met and additional arguments must be found that insufficiencies in the ML model remain within the bounds required for them not to lead to safety-relevant failures. In many cases the evaluation of these properties will require more targeted tests. However, some approaches based on introspection (e.g., white-box coverage and analysis of features maps of deep neural networks) and the formal verification of certain properties have already been proposed. Properties of the models that should be considered depend on the requirements allocated to the ML system and ML approach itself. The requirements allocated to the ML model and their derived validation targets may also depend on the capabilities of post-processing techniques to eliminate certain classes of errors. The properties investigated must be directly measurable at the level of the ML model itself and can include:

\begin{itemize}
    \item \textit{Error distribution: }This property requires an analysis of error rates and their distribution over the input space. For example, sporadic errors that occur frequently but distributed across the input space, may be filtered by post-processing techniques. However, consecutive errors may be harder to detect. Here it may also be necessary to differentiate between different classes of errors such as false positives and false negatives. False negatives may be more difficult to detect in post-processing than false positives, yet may have a more critical impact on the safety goal. An optimisation of the model towards the minimisation of false negatives, whilst allowing for a certain number of sporadic false positives may therefore be preferential to an equal distribution of both error classes. This highlights the limited applicability of typically used benchmark metrics for evaluating overall ML model performance based on precision and recall such as area under curve (AUC) \cite{davis2006relationship}. 
    
    \item \textit{Semantic analysis of triggering conditions: }This property relates to the ability of the ML model to correctly process inputs from all relevant equivalence classes of the input domain. This may include ensuring a systematic coverage of an ontological definition of the operation domain as well as a focused search for corner cases, for example based on perturbations of inputs to adjust various properties that could trigger errors \cite{hutchison2018robustness,pezzementi2018putting}. 
    
    \item  \textit{Robustness: }This property measures the resilience of an ML system to slight changes in the inputs. The origin of these disturbances can either be natural, e.g. sensor noise, poor weather conditions or corruptions \cite{hendrycks2019benchmarking}, or intentional manipulations to fool the algorithm, called adversarial attacks \cite{szegedy2013intriguing}. Robustness is typically evaluated using local robustness metrics defined in terms of the amount of perturbation around a particular data point that can be tolerated before provoking an incorrect output \cite{agarwal2018explainable,fawzi2018analysis}. The local robustness can be quantified by two bounds: the lower bound is the minimum distance around an input where no adversarial example can be found, whereas the upper bound is the maximum size of a perturbation for which no adversarial examples can be constructed \cite{serban2020adversarial}.
    
    \item \textit{Introspective measures: }Additional properties of the ML model can be used to further understand insufficiencies and performance limitations. These can include measures that increase the explainability of the calculations or the coverage achieved by certain tests \cite{huang2020survey}. However, this class of methods may not necessarily directly measure properties that could lead to errors and thus safety-related failures in the ML system but may complement or increase confidence in other forms of analysis, such as those described above. E.g. use of heat-mapping for analysing semantic causes for failed inputs. 
\end{itemize}

The targeted evaluation of specific properties of the ML model leads to the definition of a refined set of validation targets for the ML model, and a set of $1..n$ evidence functions $E_{k}$ for measuring each of these properties, which may also include the direct measurement of failure rates of the ML system as described in category 1. Equation \ref{Eq_Combined_Insufficiencies} defines the condition that if the evidence used to evaluate all relevant properties of the model can be demonstrated to meet their respective validation targets, then it can be \textit{inferred} that the ML system will meet its acceptance criteria. The challenge in safety assurance is to find a set of evidences such that the ``leap of faith'' required to infer the actual performance of the system, illustrated here by the inductive implication $\Rightarrow_{ind}$, is as small as possible.

\begin{equation}\label{Eq_Combined_Insufficiencies}
 \begin{array}{c}
 	((S_1 \subseteq I \land \forall j \in S_1.A(j) \Rightarrow E_{1}(S_1, M_M) \geq VT_{1}) \land \\
 	\cdots \\
  	(S_n \subseteq I \land \forall j \in S_n.A(j) \Rightarrow E_{n}(S_n, M_M) \geq VT_{n})) \\ 
  	\Rightarrow_{ind} 1 - \lambda_{MS}\geq AC
 \end{array}
\end{equation}

As per the causal model presented in Section \ref{Sec_Causal_Model}, the properties investigated in the model must correspond to insufficiencies and errors that have the potential for safety-relevant failures. A clear correlation between the properties under investigation, the effectiveness of the methods to accurately measure these properties and the residual safety-relevant failures must also be argued. 

\subsection{Category 3: Operation-time measures} \label{Sub_Operation_time}
For many applications, it will not be possible to reduce insufficiencies in the ML model to an extent sufficient to guarantee that residual failures in the ML system are tolerably low (as defined in Equation \ref{Eq_Combined_Insufficiencies}). In such cases, \textit{operation-time} measures can mitigate the risk associated with such residual errors to a tolerable extent. The effectiveness of such measures can be determined by the proportion of potentially hazardous residual errors of the model mitigated during operation of the system. In line with the causal model laid out in this paper, the effectiveness of the operation-time measures will be argued for each class of potentially hazardous insufficiency and error identified during the safety analysis described in Section \ref{Sec_Causal_Model}. A number of operation-time measures would typically be combined based on the requirements of the system and the results of the safety analysis and can include the following forms:

\begin{itemize}
    \item \textit{Extension of the ML model:} This category of measures involves extending the ML model to provide information about insufficiencies to consumers of its outputs in order to  predict when errors are likely to occur. Examples of which include approaches to evaluate prediction certainty at run time \cite{henne2020benchmarking} and out-of-distribution detection based either on the inputs to the ML or through introspection of model properties such as anomalous feature activation \cite{schorn2020facer}.
    
    \item \textit{Analysis of errors in the ML model outputs:} These measures treat the ML model as a black box and observe properties of the outputs to detect erroneous results. These may include plausibility checks for some types of errors (for example, feasible rates of change of pedestrians over time). 
    
    \item \textit{Redundant calculation of the target function: }Some types of errors, for example, predicting the presence of false negatives may require more complex measures involving alternative algorithms and inputs (for example, based on an analysis of radar data to identify potential areas of false negatives within an image). 
    
    \item \textit{Measures external to the ML system:} Within the causal model described in this paper, such measures would lead to a refinement of the safety requirements and acceptance criteria to allow for a more permissive level of performance in the ML system and model. For example, reducing vehicle speed in certain weather conditions which could have a beneficial impact on the occurrence and impact of certain classes of errors.
\end{itemize}

The effectiveness of such measures can be determined based on the proportion of potentially hazardous residual errors mitigated during operation. The effectiveness of the measures will need to be argued for each class of insufficiency or error to be detected. An acceptable level of residual failures will therefore likely only be achieved by a judicious combination of operation-time measures based on a thorough understanding of the bounds of insufficiencies and errors in the ML model. This leads to the  following extension of equation \ref{Eq_Combined_Insufficiencies}:

\begin{equation}\label{Eq_Operationtime}
 \begin{array}{c}
 	((S_1 \subseteq I \land \forall j \in S_1.A(j) \Rightarrow E_{1}(S_1, M_M) \geq VT_{1}) \land (OT_1(S_1, M_M) \geq DC_1) \land \\
 	\cdots \\
  	(S_n \subseteq I \land \forall j \in S_n.A(j) \Rightarrow E_{n}(S_n, M_M) \geq VT_{n}) \land( OT_n(S_n, M_M) \geq DC_n)) \\ 
  	\Rightarrow_{ind} 1 - \lambda_{MS}\geq AC
 \end{array}
\end{equation}

Where the function $OT_k$ returns the proportion of insufficiencies or errors in the ML model of class $k$ that are detected by the operation-time measures (diagnostic coverage) and $DC_k$ is the required diagnostic coverage related to the property $k$ for the overall acceptance criteria $AC$ to be met. The target values for $VT_k$ and $DC_k$ will depend on the requirements on the ML system and the nature of each property $k$.

\subsection{Category 4: Design-time measures} \label{Sub_Design_time}
If the evidence collected within categories 1..3 do not lead to a sufficiently convincing argument that the acceptance criteria are met, then additional design-time measures may be required to further reduce the probability of insufficiencies in the ML model. Such measures would be informed by the evaluation of the evidence collected in categories 1..3 and could include the following:

\begin{itemize}
    \item \textit{(Re-)Definition of the operating and system context:} These measures, related to operation-time measures external to the ML system (e.g. limiting operation to specific scenarios or the introduction of addition redundancy measures), can lead to a refined, specification of the safety requirements and acceptance criteria and validation targets allocated to the ML system.
 
    \item \textit{Selection of the most appropriate ML-technique: } Causes of insufficiencies could also include the inherent inability of the ML model to accurately represent the complex target function based on the chosen ML approach, architecture or hyper-parameters. Justification that the chosen ML approach and architecture has the inherent potential to fulfill its allocated safety requirements and acceptance criteria will therefore be an essential part of the safety argument.

    \item \textit{Criteria on training data: } These measures refine the set of criteria used to select suitable training data. This may be necessary based on the semantic analysis of triggering conditions or other causes of errors that result from over-fitting (causing high variance) and under-fitting (causing high bias). Whereby models can suffer from over-fitting for some parts of the ODD and under-fitting in others requiring bias-variance trade-offs to be made.
    
    \item \textit{Optimization of the ML system architecture and ML model: } This measure optimizes the design, e.g. choice of hyperparameters) of the ML model itself in order to reduce insufficiencies but can also include the extension of operation-time mechanisms to better mitigate residual errors.
\end{itemize}

The effectiveness of the chosen design-time measures will be validated by evidence collected in categories 1 and 2 leading also to a possible refinement of the operation-time measures (category 3) required to minimise residual safety-related failures. This cycle, illustrated in Figure \ref{Fig_Assurance_Framework} will need to be repeated until an acceptable level risk has been achieved  based on the quantitative results of the analyses as well as a qualitative evaluation of their effectiveness.

\section{Example Analysis}
This section demonstrates the concepts described within this paper within the context of a pedestrian recognition function for automated driving \cite{gauerhof2018structuring}. The acceptance criteria for the function can be expressed as follows:

\begin{itemize}
    \item Each pedestrian within the \textit{critical range} is correctly detected within any sequence of \textit{N images} with a \textit{true positive rate}, \textit{vertical} and \textit{horizontal deviation} from ground truth sufficient to avoid collisions.
\end{itemize}

Where parameters which must be quantitatively defined specific to the application and context are highlighted in \textit{italics}. The system is considered to fail whenever the above condition is not met. For example, the magnitude or rate of errors exceeds the bounds defined by the parameters.

Inductive analyses similar to Failure Modes and Effects Analysis (FMEA) 
can be used to analyse the various types of error classes and underlying insufficiencies in the model that could lead to the violation of a safety goal and their underlying causes. Table \ref{table_causal_analysis} includes an excerpt of such an analysis, including underlying causes and counter-measures applied either during design-time or during operation. Alternatively, deductive forms of safety analysis could include evaluating specific safety-related failures discovered during validation to determine the underlying causes and improvements to the ML system required to prevent their occurrence in future.

In practice, a combination of inductive and deductive analysis is required in order to understand the factors that influence the safety-related properties of the ML system and to successively develop a set of measures sufficient to demonstrate that the acceptance criteria are met. The effectiveness of each measure should be evaluated according to a set of metrics which in turn can be compared to validation targets for each property. The selection of the threshold values for the validation targets will depend not only on the properties of the system-level acceptance criteria but also on the correlation between the error classes, their causes and the potential of the design and operation-time measures. For example, for some classes of errors, such as false negatives, no effective operation-time measures may exist, thus increasing the demands on design-time measures to mitigate against this error class. Likewise, for false positives, post-processing filtering mechanisms may exist with a high diagnostic coverage, thus allowing for trade-offs to the benefit of reducing false negatives during design to be made. Thus, the safety assurance task can also be seen as an exercise in the global optimization of a number of potentially conflicting properties of the trained ML model.

\begin{table*}[ht]
   \centering
    \begin{tiny}
        \begin{tabularx}{\textwidth}{|X|X|X|X|X|X|X|}
        \hline
         \textbf{Error class} & \textbf{Insufficiency} & \textbf{Cause} & \textbf{Design-time measure} & \textbf{Metric} & \textbf{Operation-time measure} & \textbf{Metric}\\ 
         \hline
         Incorrect classification  & Lack of generalization & Under-specification, scalable oversight  & Balanced training set & Coverage of ODD model & N/A & N/A\\
         \hline
         Incorrect classification & Unreliable confidence values & Over-confidence due to uncalibrated soft-max values & Temperature scaling & Remaining Error Rate, Remaining Accuracy Rate & N/A & N/A \\
         \hline
         ... & ... & ... & ... & ... & ... & ...\\
         \hline
         False negatives & Lack of robustness & Instability of DNNs for minor changes to the inputs & Adversarial training & Adversarial and perturbation robustness & Robustness certificates & Certifiable perturbation strength  \\
         \hline
         Sequence of false negatives & Lack of generalization & Under-specification, scalable oversight & Balanced training set & Coverage of ODD model & Comparison with other sensor data & Diagnostic coverage \\
         \hline
         ... & ... & ... & ... & ... & ... & ...\\
         \hline
         False positives & Clever Hans effect & Spurious correlations in the training data & Diversified training set & Conceptual disentanglement & Plausibility checks & Diagnostic coverage \\
         \hline
         False positives & Lack of generalization & Distributional shift & N/A & N/A & Out of distribution detection & Diagnostic coverage \\
         \hline
         ... & ... & ... & ... & ... & ... & ...\\
         \hline
        \end{tabularx}
    \end{tiny}
    \caption{Excerpt of the analysis of error classes, insufficiencies, causes, measures and metrics for a pedestrian detection function}
    \label{table_causal_analysis}
\end{table*}

\section{Conclusions and future work} \label{Sec_Conclusions}
Safety assurance of ML-based systems must consider uncertainties within the model and its environment leading to the need for an evaluation of performance limitations and insufficiencies. This paper has described a causal model for reasoning about the safety of the intended functionality (SOTIF) with respect to machine learning systems. This led to the definition of 4 categories of safety evidences that, when combined, could lead to an assurance case based on a quantitative and qualitative arguments that the acceptance criteria on the system are met. The definitions were supported by formalisms that illustrated the \textit{inductive reasoning} approach to gathering convincing evidence for the fulfilment of the acceptance criteria. Each category of evidence in itself would typically not be sufficient as a basis of a convincing argument, a holistic consideration is therefore required, where assumptions in each category are strengthened by the results of the others. 

Previous work in his area (see Section \ref{Sec_Related_Work}), though scientifically valid and relevant, has not yet to lead to a convincing consensus regarding the potential of ML-based systems to fulfill the safety requirements of modern automated systems. The causal model described here leads to two main observations why this may be the case. Descriptions provided by the safety community of processes and structured assurance cases for machine learning have yet to define detailed criteria on the evidences required to support these arguments. Whilst detailed work on safety-relevant metrics and methods, typically originating from the machine learning community, have focused on the optimization of individual properties of the models leading to a fragmented landscape of metrics and a lack of clarity whether or not the methods could eventually lead to an acceptable level of safety. The formalisms provided in this paper should provide inspiration for more rigorous development and evaluation of ML-related safety measures and metrics in future.

In order to make progress in this field we therefore see the need for the safety and machine learning communities to work more strongly together and in particular to address the following challenges:

\begin{itemize}
    \item Machine learning itself is based on statistical modelling techniques, whilst the properties of the environment (triggering conditions) that can lead to failures can often only be reasoned about in a probabilistic manner due to the complexity of the operating domain and lack of sufficient environmental models. It should therefore come as no surprise that, unlike previous approaches for traditional software-based systems, the safety assurance of machine learning will require statistical arguments regarding the residual failure rates of the system.
    
    \item To support the validity of these statistical arguments, there is a need to understand the causality between the underlying causes of insufficiencies and how these propagate to errors in the outputs of the ML model and eventual failures of the ML system to meet its safety goals. These causalities will need to be identified based on qualitative engineering judgements but also empirically investigated statistical correlations between various measurable properties of the inputs, the models and safety evidence. A structured approach to safety analysis as presented in Section \ref{Sec_Causal_Model} would support such evaluations for various classes of insufficiencies. 

    \item Due to the growing range of ML techniques and applications, there is likely to be no ``one size fits all'' set of methods and metrics that can be used to form such safety arguments. However, over time, a set of proven methods and metrics for certain applications and ML techniques may emerge. This will require a confirmation of the effectiveness of the methods and an iterative approach to forming the assurance case until a qualitative consensus of safety can be reached. In particular, we see a responsibility on the research community when presenting methods for ML safety to demonstrate their effectiveness at both evaluating specific properties of the ML model and reducing the overall rate of safety-related failures.
\end{itemize}

Future work will apply the approaches described here to applications of ML safety within various domains in order to refine the framework and provide convincing contributions to emerging safety standards and assessment approaches in the form of concrete methods for each of the 4 contribution categories outline here. This will also also involve an evaluation of the effectiveness of currently proposed methods and metrics against the criteria for effective evidence outlined in Section \ref{Sec_Categories}. Furthermore, work is currently underway on the first generation of standards for the use of ML in safety-critical cyber-physical systems (see ISO PAS 8800 and ISO/IEC TR 5469 as two examples). We propose that these standards refer to a causal model of safety and conditions for effective arguments as outlined in this paper.

\subsubsection*{Acknowledgements}
This work was funded by the Bavarian Ministry for Economic Affairs, Regional Development and Energy as part of a project to support the thematic development of the Institute for Cognitive Systems and within the Intel Collaborative Research Institute Safe Automated Vehicles. I would like to thank Chih-Hong Cheng and Karsten Roscher for the valuable support and suggestions in preparing this paper.

\bibliographystyle{splncs04}
\bibliography{Foundations.bib}

\end{document}